\documentclass[aps,prr,twocolumn,superscriptaddress]{revtex4-2}

\usepackage[english]{babel}
\usepackage[utf8]{inputenc}
\usepackage{graphicx}
\usepackage[unicode=true,pdfusetitle,
 bookmarks=true,bookmarksnumbered=false,bookmarksopen=false,
 breaklinks=false,pdfborder={0 0 0},pdfborderstyle={},backref=false,colorlinks=true]{hyperref}
 \hypersetup{
 citecolor=blue,
 urlcolor=blue,
 linkcolor=blue}

\usepackage{latexsym,amsmath,verbatim}
\usepackage{amssymb} 
\usepackage{amsfonts}
\usepackage{array}
\usepackage{stackrel}
\usepackage{bm}
\usepackage{nicefrac}
\usepackage{float}

\makeatletter
\AtBeginDocument{\let\LS@rot\@undefined}
\makeatother

\newcommand{\new}[1]{{#1}}

\begin{document}

\title{Laplacian paths in complex networks: information core emerges from entropic transitions}


\author{Pablo Villegas}
\email[pablo.villegas@cref.it, guido.caldarelli@unive.it]{}
\affiliation{Networks Unit, IMT School for Advanced Studies Lucca, Piazza San Francesco 19, 50100, Lucca, Italy}
\affiliation{``Enrico Fermi" Research Center (CREF), Via Panisperna 89A, 00184 - Rome, Italy}
\author{Andrea Gabrielli}
\affiliation{Dipartimento di Ingegneria, Università Roma Tre, 00146, Rome, Italy}
\affiliation{``Enrico Fermi" Research Center (CREF), Via Panisperna 89A, 00184 - Rome, Italy}
\affiliation{Institute for Complex Systems, Consiglio Nazionale delle Ricerche, UoS Sapienza, 00185 Rome, Italy}
\author{Francesca Santucci}
\affiliation{Networks Unit, IMT School for Advanced Studies Lucca, Piazza San Francesco 19, 50100, Lucca, Italy}
\author{Guido Caldarelli}
\email[pablo.villegas@cref.it, guido.caldarelli@unive.it]{}
\affiliation{Department of Molecular Sciences and Nanosystems, Ca' Foscari University of Venice, 30172 Venice, Italy}
\affiliation{European Centre for Living Technology, 30124 Venice, Italy}
\affiliation{London Institute for Mathematical Sciences, W1K2XF London, United Kingdom}
\author{Tommaso Gili}
\affiliation{Networks Unit, IMT School for Advanced Studies Lucca, Piazza San Francesco 19, 50100, Lucca, Italy}

\begin{abstract}
Complex networks usually exhibit a rich architecture organized over multiple intertwined scales. Information pathways are expected to pervade these scales reflecting structural insights that are not manifest from analyses of the network topology. Moreover, small-world effects correlate with the different network hierarchies complicating the identification of coexisting mesoscopic structures and functional cores. We present a communicability analysis of effective information pathways throughout complex networks based on information diffusion to shed further light on these issues. We employ a variety of brand-new theoretical techniques allowing for: (i) bring the theoretical framework to quantify the probability of information diffusion among nodes, (ii) identify critical scales and structures of complex networks regardless of their intrinsic properties, and (iii) demonstrate their dynamical relevance in synchronization phenomena. By combining these ideas, we evidence how the information flow on complex networks unravels different resolution scales. Using computational techniques, we focus on entropic transitions, uncovering a generic mesoscale object, the \emph{information core}, and controlling information processing in complex networks. Altogether, this study sheds much light on allowing new theoretical techniques paving the way to introduce future renormalization group approaches based on diffusion distances.
\end{abstract}

\maketitle

Transmission and processing of information in complex networks are functions of the underlying spatial graph structure determining the paths along which the information flows. Such paths strongly depend on the spatial resolution at which the dynamical processes operate \cite{Masuda2017, Cimini2019}. We can even say that a particular flow --strongly conditioned by the underlying topology--  is directly linked to the network scales through the network ``communicability'', i.e., how a perturbation on one node of the network is ``felt'' by the rest of the nodes with different intensities \cite{Estrada2012}. We need, therefore, to consider three powerful concepts to shed light on the interaction between information flow and graph structure. 

The first one is Shannon's entropy \cite{shannon1948} that is related to the ``amount of information'' contained in a probability distribution allowing, for example, to find the probability of the available microstates of the classical statistical ensembles \cite{Jaynes,Pathria}. In the network community, it has proven to be essential to reveal the time scale dependences in neural systems \cite{borst1999, strong1998}, \new{to characterize network ensembles \cite{Ginestra1}} or to unravel different mesoscopic structures through random-walk diffusion techniques \cite{Esquivel2011, Rosvall2008, Rosvall2014}. In particular, recent pioneering works have proposed a set of information-theoretic tools formalizing an entropy measure for complex networks both in simple graphs and multilayer networks \cite{Domenico2016, Domenico2015,ghavasieh2022}. It is not, however, intention of this paper to shed light on the profound debate of what information is \cite{Adami2016}, but on how information is stored in a network and what mesoscopic units play an essential role in its processing and transmission. Hence, its profound meaning and implications remain a crucial question to be answered.

Heterogeneous scale-dependent structures have been proposed to emerge as an optimal solution when resources are scarce, and there is some cost involved in forming connections between nodes \cite{Csermely2013}. In particular, core structures are expected to play a crucial role in supporting integrated network function in the brain and genetic networks \cite{Bullmore2012, Fornito2016}. Also, rich-club structures enrich the functional repertoire over and above the effects produced by scale-free type topology \cite{Senden2014}. In contrast, core-periphery structures foster the existence of a central integrative functional core \cite{yizhar2011, Gollo2015} (see \cite{Fornito2016} for a comprehensive review). A fundamental open challenge involves characterizing mesoscale objects such as giant components and functional cores in terms of diffusion geometry \cite{Boguna2021}.
This leads us to consider the second ingredient for our analysis: a mechanism of diffusion in the system, able to capture the network properties such as small-worldness, degree
heterogeneity, or clustering. The usual mechanism of diffusion in the case of graphs takes the form of the Laplacian matrix \cite{Ginestra2020, Masuda2017}, which encodes the heterogeneity properties of the network through its spectrum of eigenvalues and corresponding eigenvectors. Note that diffusion is a fundamental ingredient of most of the studied dynamical processes on, for instance, networks synchronization \cite{Ginestra2019}.

The third and final ingredient we need is a theoretical tool to characterize the information diffusion and related dynamics in any heterogeneous network at different scales, \new{i.e., a field theory of information dynamics in heterogeneous structures}. In regular structures (i.e., regular lattices), the renormalization group (RG) is the fundamental theory that permits the accurate analysis of static and dynamical statistical physics models at different scales providing an elegant and precise theory of criticality. It allows to connect --via the scaling hypothesis-- extremely varied spatiotemporal scales and to understand the fundamental issues of scale invariance \cite{Binney, Amit, Kardar}. Unfortunately, due to the strong topological heterogeneity, its complex network counterpart still presents serious issues. All the current approaches suffer from several limitations \cite{Garlaschelli} (e.g., assumption of specific topological properties \cite{Garcia2018}, limited iterability in networks with small-world effects or irreducibility to the ordinary scheme for regular lattices). Still, Zheng et al. \cite{Zheng2020} performed an RG-inspired approach for the Human Connectome by studying zoomed-out layers showing that they remain self-similar under specific coarse-graining transformations of nodes and connectivity \cite{Zheng2020, Garcia2018}. \new{To develop an RG scheme in heterogeneous networks, it is, therefore, crucial to making progress in this direction: to extend recent approaches which have provided the basis to develop a statistical field theory of information dynamics on complex networks \cite{Domenico2016,ghavasieh2022}, on the grounds of information fluxes between nodes \cite{ghavasieh2020}.}


\new{The heterogeneous topology} of a network, characterized by peculiar structures, determines how the information flow at different scales on top of the network \cite{Bullmore2012}. Hence, in-depth knowledge of the dependence on structural network properties of the information diffusion is essential to interpret collective network phenomena from a dynamical point of view. For instance, the interspersed nature of multiple pronounced resolution scales suggested the existence of stretched criticality regions \cite{NatComm} both in the activity spreading dynamics \cite{NatComm, MAM2010, MAM2012} and in the appearance of broad frustrated synchronization regimes \cite{Villegas1, Villegas2}. Moreover, genuine scale-invariant networks, such as the  Barab\'asi-Albert one \cite{Barabasi1999}, show no sign of epidemic threshold \cite{Pastor2010} even though they still present an unforeseen non-vanishing synchronization transition point \cite{yamir2004, Arenas2008} with no reasonable explanation up to now.

In this work, we develop a statistical physics framework that, grounding on the concept of Shannon entropy, permits us to study the fundamental paths along which information is transmitted throughout complex networks. 
In particular, we first introduce the Laplacian network propagator at different times and the spectral entropy through the measure defined by its spectrum. We then study the variations of this entropy as a function of the diffusion time, revealing essential substructures and modules at different resolution scales. More precisely, we show that such entropy acts as an order parameter for structural phase transitions, and its derivative plays the role of specific heat. Indeed in networks characterized by a complex hierarchical organization of scales (e.g., the Human Connectome), the different resolution scales at which such specific heat shows pronounced peaks identify the characteristic intrinsic network scales. Moreover, as we explicitly demonstrate, these are precisely the fundamental scales uncovering different functional cores playing a crucial role in network dynamical processes such as, for instance, synchronization.



\section{Results}
\subsection{Canonical formulation}
Information diffusion in complex networks rules as set out by the Laplacian matrix $\hat L$ \cite{Newman,Masuda2017}, defined for undirected networks as $L_{ij}=\delta_{ij}\stackrel[k]{}{\sum}A_{ik}-A_{ij}$, where $A$ stands for the network's adjacency matrix \footnote{Under the only condition of being a connected network, to ensure that all the nodes can explore all the accessible points in phase space, thus fulfilling the ergodic hypothesis.}, and $\delta$ is the Kronecker delta function. The Laplacian thus regulates the evolution of information of a given initial specific state of the network, $\textbf{s}(0)$, which will evolve with time as $\textbf{s}(\tau)=e^{-\tau \hat L}\textbf{s}(0)$. The \emph{network propagator},  $\hat K=e^{-\tau \hat L}$, represents the discrete counterpart of the path-integral formulation of general diffusion processes \cite{Feynman}, and each matrix element $K_{ij}$ substantially accounts for the sum of diffusion trajectories along all possible paths connecting nodes i and j up to a temporal scale $\tau$ \cite{moretti2019,Masuda2017}.

In terms of the network propagator, $\hat K$, let us now define the operator \cite{Domenico2016}, 
\begin{equation}
 \mathbf{\rho(\tau)}= \frac{\hat K}{\mathrm{Tr}(\hat K)}=\frac{e^{-\tau \hat L}}{Tr(e^{-\tau \hat L})},
 \label{EvolMat}
\end{equation}
whose eigenvalues $\mu_i(\tau)$ with $i=1,2,...,N$ are simply related to the eigenvalues $\lambda_i$ of $\hat L$ by
\begin{equation}
\mu_i(\tau)=\frac{e^{-\lambda_i\tau}}{\sum_j e^{-\lambda_j\tau}}\,.
\label{mu}
\end{equation}
Note that the generic eigenvalue $0<\mu_i(\tau)\le 1$ gives the relative weight of the corresponding Laplacian eigenvector in the eigenvector decomposition of the network state ${\bf s}(t)$. By the properties of the Laplacian for a simple connected graph, we have that all eigenvalues of $\hat L$ are positive with the only exception of the minimal one $\lambda_{min}=0$ whose corresponding eigenvector is uniform. Consequently, through the measure  given by Eq.~\eqref{mu} we can define the Shannon entropy --at time $\tau>0$ -- (as recently proposed in \cite{Domenico2016,ghavasieh2020}) 
\begin{equation}
 S[\rho(\tau)]=-\frac{1}{\log(N)}\stackrel[i=1]{N}{\sum}\mu_{i}(\tau)\log\mu_{i}(\tau),
 \label{S-rho}
\end{equation}
where he normalization coefficient $1/\log(N)$ makes the entropy $S[\rho(\tau)]\in [0,1]$. Indeed its maximum value is obtained in the case of $N$ identical eigenvalues $\mu_i(\tau)=1/N$ (for $\tau=0$), which describes the most trivially heterogeneous network composed by $N$ isolated (i.e., independent) nodes. As shown below, this will allow us to consider it as a potential order parameter for the study of \emph{entropic phase transitions} (or information propagation transitions, i.e., diffusion) over the network. We state here a heuristic explanation of such quantity: let us assume to start the dynamics with a generic heterogeneous state, $\mathbf{s}(0)$, having non-null components along all eigenvectors of $\hat L$. The more heterogeneous $\mathbf{s}(0)$ is in terms of its decomposition along the eigenvectors of $\hat L$, the larger the information encoded in the network. In this sense, $S[\rho(\tau)]$ can be seen as a measure of the residual information still encoded in the evolved state $\mathbf{s}(\tau)$. Note that this formulation does not consider any information about the initial state of the nodes and only draws on the properties of the network's propagator that encodes the information streams or information flow between nodes $i$ and $j$ \cite{ghavasieh2020}. We stress the specific application considering $\tau=-1$, the so-called \emph{Laplacian Estrada index} of a network initially proposed to quantify the degree of folding of long-chain molecules \cite{Estrada2000}, which also provides a centrality measure of the network \cite{Estrada2012, Bamdad2010}.

\subsection{Entropic phase transitions}

\begin{figure}[hbtp]
 \includegraphics[width=1.0\columnwidth]{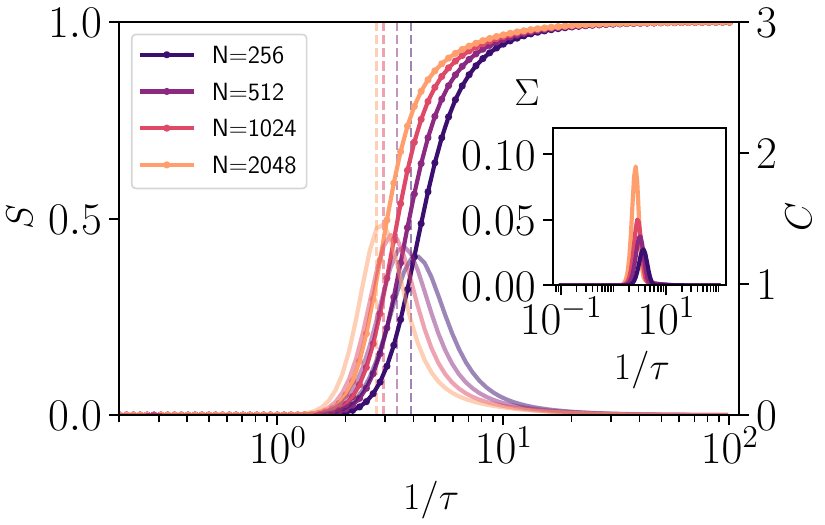}
 \caption{Network average of the entropy $S[\rho(\tau)]$, versus the inverse of time evolution $\nicefrac{1}{\tau}$, for an Erd\"os-Renyi network of $\langle k \rangle=30$ and different system sizes (see legend). A critical point ($\tau_C$) separates the segregated phase from the integrated one. The specific heat, $C$, presents a peak just at this critical value. Inset: Variance of the entropy, \new{averaged in the graph ensemble}, and multiplied by $N$; $\Sigma=\sigma^2_S N$. The point of maximal variability coincides with the point of maximal slope in $S[\rho(\tau)]$ for all network sizes $N$ (dashed lines in the main figure). All curves have been averaged over $10^2$ realizations.}\label{Entropies}
\end{figure}

We perform an extensive computational study of the entropy $S[\rho(\tau)]$ of different network structures. Our results reveal the existence of {\em entropic} second-order phase transitions accounting for the information diffusion and processing throughout the network (see, e.g., Fig.\ref{Entropies} for the case of an Erd\"os-Renyi network). 
By increasing the diffusion time $\tau$ from $0$ to $\infty$,  $S[\rho(\tau)]$ decreases from $1$ (\emph{segregated} and heterogeneous phase -- the information diffuses from single nodes only to the local neighborhood) to $0$ (\emph{integrated} and homogeneous phase -- the information has spread all over the network) in all connected simple graphs. The derivative of the entropy concerning the (logarithm of the) diffusion time $\tau$
\begin{equation}
    C=-\frac{dS}{d(\log \tau)},
\end{equation}
is a detector of transition points corresponding to the intrinsic characteristic diffusion scales of the network. Indeed, a pronounced peak of C defines $\tau=\tau_C$ and reveals a strong deceleration of the information diffusion, separating regions of the network with strong diffusion from the rest of the network where the diffusion slows down. 
To clarify this point, we can use the analogy with thermodynamic systems. More precisely, since for a simple graph $\hat L$ is a Hermitian matrix, we can see Eq.~\eqref{EvolMat} as a canonical density operator of statistical physics in which $\hat L$ plays the role of the Hamiltonian operator and $\tau$ the role of the inverse temperature \cite{Binney, Pathria, Greiner}. In this sense, $S[\rho(\tau)]$ corresponds to the canonical Von Neumann entropy \cite{Domenico2016} and its derivative concerning $\log \tau$ is the specific heat of the system. A sharp maximum of this quantity, which diverges in the thermodynamic limit, is a fingerprint of a second-order phase transition in statistical physics. Moreover, thanks to this analogy, we can use the thermal fluctuation-dissipation theorem \cite{Marro, Moloney} connecting the specific heat to the entropy fluctuations saying that $C$ is proportional to $\sigma^2_S=\langle S^2 \rangle - \langle S \rangle^2$, \new{where $\langle \cdot \rangle$ indicates fluctuations over the graph ensemble}. In particular, we expect that $\sigma^2_S$, over many independent network realizations, scales as $1/N$ where $N$ is the number of nodes of the network (as a direct application of the central limit theorem \cite{Gardiner}). The inset of Fig.\ref{Entropies} shows the scaled variance of the entropy, $\Sigma = N\sigma^2_S$, which exhibits a pronounced peak at the transition point, revealing anomalous scaling as expected at true criticality. The combination of these quantities ($S[\rho(\tau)]$, $C$ and $\Sigma$) allows us to affirm the existence of a bonafide second-order phase transition.


\begin{figure*}[hbtp]
 \includegraphics[width=2.0\columnwidth]{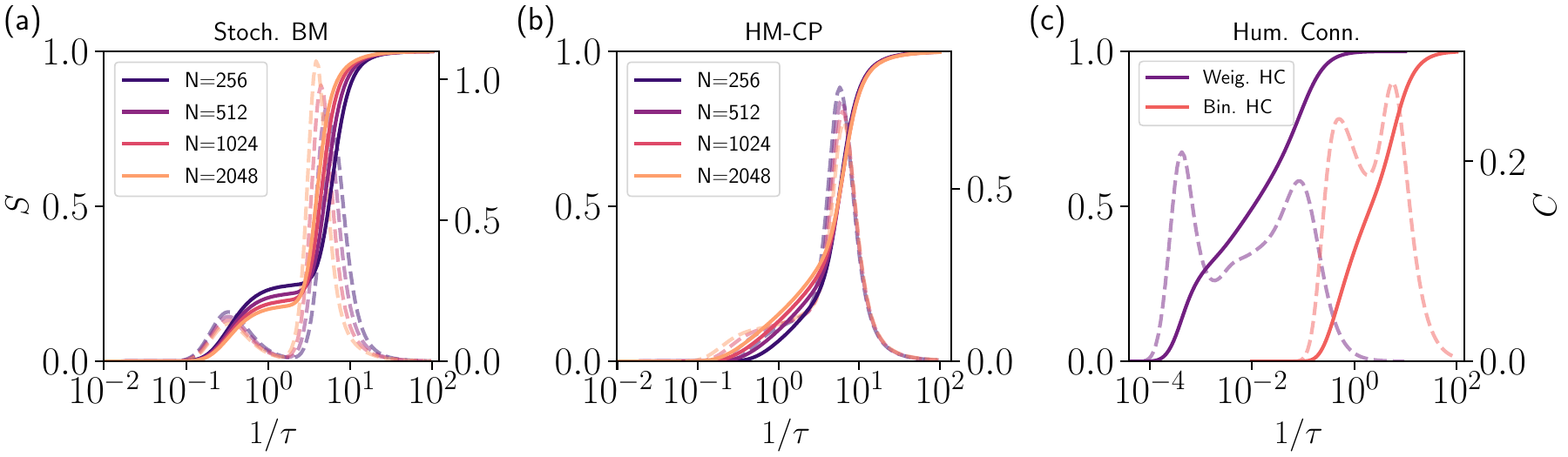}
 \caption{Network average of the entropy parameter $S$, versus the inverse of time evolution $\nicefrac{1}{\tau}$ for: (a) Stochastic block model \new{(SBM)} for different network sizes (see legend) constituted by four equal interconnected modules $p=128/N$ and interconnectivity probability $q=1/N$ \new{(i.e., with $\langle k \rangle=32$)}. Two peaks in the derivative of the order parameter $C$ indicate diverse critical points, and a broad region separates the segregated phase from the integrated one. \new{The peak at short times happens for similar diffusion times to the ER network, because of the similarity between both networks at a local scale.} (b) Hierarchic modular network with core-periphery structure (HC-CP). We consider a set of basal nodes with $N_b = 25$ units and connectivity per node $k_0 = 12$ \new{(the mean connectivity ranges from $\langle k \rangle \simeq 37$ for $N=256$ to $\langle k \rangle \simeq 52$ for $N=2048$)}. (c) Human connectome network \new{($\langle k \rangle \simeq38$ for the binarized case)}.  All curves for the SBM and the HM-CP have been averaged over $10^2$ realizations.}\label{Networks}
\end{figure*}

Once described the expected system phases, we now discuss different underlying network structures and their emergent effects. For all connected networks, the averaged entropy, $S$, shifts from a distinctively zero value (i.e., full integration of information) to a completely isolated set of nodes but featuring an entirely different transient phenomenology. In particular, homogeneous networks, as the paradigmatic example of the Erd\"os-Renyi network, show a second-order phase transition depending on $\tau$ values (as shown in Fig.\ref{Entropies}), capturing the information flowing from small subsets of nodes at the very beginning, to an effective network acting as a whole for considerable times. A completely different phenomenology emerges when analyzing networks with further complexity and interspersed scales as, e.g., stochastic block models (SBMs, see Appendix \ref{CorePAp}). SBMs are composed of $N$ nodes organized into \new{$C_r$} subsets or communities, with different intracommunity and intercommunity connection probabilities, $p$ and $q$, respectively. Due to their particular community structure, SBMs present a representative two-peaked behaviour when examining $C$ (see Fig.\ref{Networks} (a)). This double-peaked structure reflects the probability of successfully integrating information within the modules and throughout the entire network, thus capturing the characteristic network scales. It is essential to point out that this kind of network --even if it says nothing new-- constitutes the most straightforward example that allows for an emergent complex dynamical compromise between segregation and integration for intermediate diffusion times.

We now consider more sophisticated multi-scale networks, hierarchical-modular random networks (HM-R), including different hierarchical levels (a sort of \emph{network of networks} built employing the algorithms proposed in \cite{Zamora2016}, see also Appendix \ref{NetBuild}). In particular, networks are created based on a nested stochastic block model in which modules are subdivided into further modules. Connections are made by selecting two nodes randomly and connecting them with a fixed probability that depends on the hierarchical scale. This type of network can be enriched by considering a more sophisticated building algorithm, i.e., replacing the random connectivity probability between modules with a preferential attachment rule that produces a core-periphery (HM-CP) structure involving central connector hubs having local and global rich-clubs (so that the degree distribution is scale-free \cite{Zamora2016}).

Figure \ref{Networks}(b) shows the results for an HM-CP network with a different number of hierarchical levels. Even if the peak at short times does not qualitatively change, regarding the SBM, we realized that the peak at large times displays a continuum non-monotonic decay, a sign of smooth information processing in ascending the nested hierarchical structure (see Fig.\ref{Networks} (b)). In turn, the comparison of simple HM-R networks and HM-CP networks reveals that a core-periphery structure allows for earlier processing of information on the networks together with a more extensive information containment for long diffusion times (see further details in Appendix \ref{CorePAp}). It is compliant with recent results indicating that a core-periphery structure fosters the emergence of broad synchronization regions with high dynamical variability \cite{Buendia2021}.

It is appropriate to mention the particular case of scale-free networks (SF) separately. They present a power-law scaling of the variance maximum as a function of the system size going to $\Sigma \to 0$ in the thermodynamic limit (see Appendix \ref{SFAp}). We must emphasize that it is driven by the maximum system's eigenvalue of the network Laplacian matrix, which is proportional to the cutoff of the degree distribution, $k_{max}$, thus scaling with the system size \cite{Arenas2008}. This leads, among other consequences, to justify the null epidemic threshold in unstructured networks \cite{boguna2003} (but yet allowing a finite set of synchronization \cite{yamir2004}, as discussed below). For our purposes, it is a symptom of inefficient network thorough information processing, as the accurate analysis of $C$ shows, unveiling a constant value among different scales, strongly related to the spectral properties of the network (see Appendix \ref{SFAp}).


Finally, we perform an accurate analysis of the existing Human Connectome network (HC, \new{the reconstruction of structural human brain networks --through neuro-imaging techniques--, which are composed of hundreds of neural regions and thousands of white-matter fibre interconnections \cite{Hagmann2008, Honey}).} Figure \ref{Networks}(c) shows \new{(for the binarized and weighted HC)} the existence of a robust multi-scale double-peaked transition indicative of two pronounced hierarchical singular scales finding evidence of scale-dependent structures (more specifically, by the emergence of well-connected modules, as discussed afterwards). We realize that the precise design of weights in the HC allows for a richer structure between the two critical network scales (i.e., enrich the complexity of the network), as the derivative of $S$ reveal, thus allowing for greater flexibility in the integration/segregation balance of information across the network.

The reader can gain insight into the emergent phenomenology in Supplemental Material 1 \cite{SI}, which contains videos showing the phase transitions for different network structures.

\begin{figure*}[hbtp]
 \includegraphics[width=2.0\columnwidth]{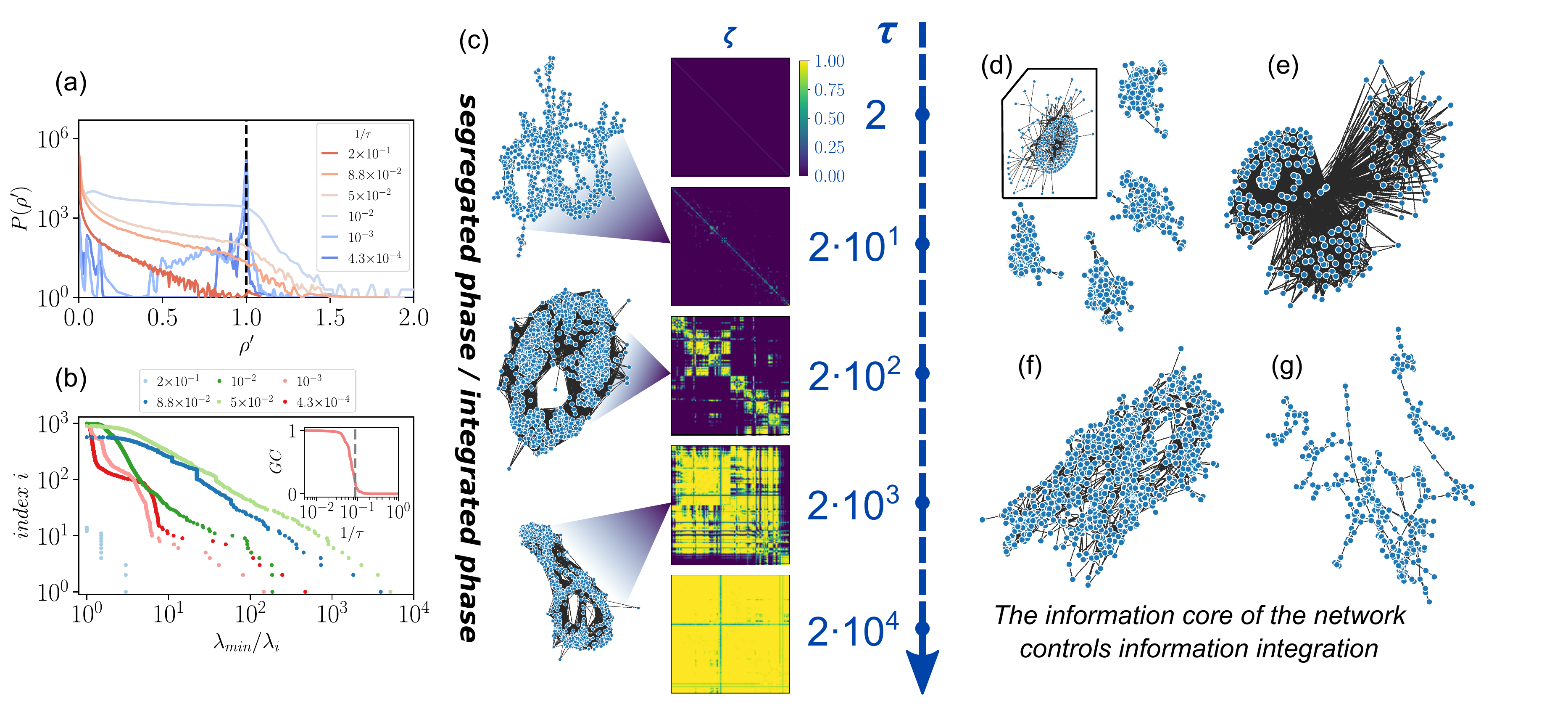}
 \caption{\textbf{Information integration substructures. (a)} Probability distribution of processed information, $P(\rho')$, for different values of the time evolution in the weighted Human Connectome network (see legend). At short times no nodes can integrate and process information (red line), generating an emergent processing information structure at intermediate times that finally converge to a delta function for large enough times (dark blue line). \textbf{(b)} Rank index i versus the normalized inverse of
the corresponding non-zero eigenvalues of the $\zeta$ Laplacian matrix for the weighted HC, at different evolution times (see legend). The decay time, associated with the different eigenvalues, reflects the different hierarchical organizations at different resolution scales. \new{Inset: Fraction of nodes which are in the giant component (GC) versus $\nicefrac{1}{\tau}$ for the weighted HC. Grey dashed line indicates the position of the C peak at short times (see purple line in Fig. \ref{Networks}c).} \textbf{(c)} Emergent structures in the weighted HC.  The time evolution, $\tau$, increases from up to down highlighting different meaningful mesoscopic scales, namely: $\tau=2 $ (segregated phase), $\tau=2\cdot10^{1}, 2\cdot10^{2}, 2\cdot10^{3}$  (intermediate phase), and $\tau=2\cdot10^4$ (integrated phase). \emph{Left-column:} The giant component of network substructures, obtained through the $\rho'$ binarized version, $\zeta$ (as explained in the main text).  \emph{Right-column:} Snapshots of typical  $\zeta$ information integration matrices; the color code represents the information integration for pair nodes as shown in the scale. The segregated phase is characterized by no information invading the system, being it confined on each node (i.e., along the diagonal, displaying the basal system scale). On the other hand, for intermediate values of $\tau$ different substructures coexist, depending on the structural complexity of the underlying network. In the segregated phase, substructures are no longer observed, and a homogeneous, 'all to all,' information integration network is observed (i.e., the network can be considered as a unique node, see also the videos in Suppl. Material \cite{SI}). \textbf{(d-g)} Information core for different networks: (d) Stochastic block model (SBM) with $N=1024$ nodes. Inset: Giant component corresponding to the particular decomposition of an individual module, (e) giant component of a hierarchic modular network with core-periphery (HM-CP) with $N=1024$ nodes and $6$ hierarchical levels, and the Human Connectome network, whether (f) binarized or (g) weighted.}\label{Connectome}
\end{figure*}

\subsection{Zooming out networks: Information core and characteristic scales}

The underlying connected substructure of information paths between the two prominent entropy peaks has profound physical implications that we discuss hereunder. To identify it, we propose a method coming from the information propagator of the network (see Eq. (\ref{EvolMat})) shed light on the most prominent network substructures.

At the very initial time, only isolated nodes are considered $\tau=0$. Instead, when time is going on, the first peak of $C$ reflects the existence of a characteristic scale below which the information diffuses rapidly and then slows down. In other words, this peak detects the first highly connected network structures (information reservoirs) where information rapidly homogenizes. We call the giant component of this set of nodes the \emph{information core} of the network. The last peak, which happens at long evolution times, takes account of full information integration all over the network (i.e., it represents the set of fully interconnected nodes). As a direct consequence, the distance between peaks is tightly interlinked with both the spectral dimension of the network and the nested hierarchical-modular structure. It takes account of the different local structures and scales of the system.

To better illustrate this effect, we analyze the evolution of the $\rho$-matrix, which encodes the \emph{effective integrated information} between each pair of nodes in the network. Observe that this matrix is originally diagonal, with all terms equal to $\rho_{ij} = \delta_{ij} /N$ at the very initial time $\tau=0$. From then on, the resulting matrix will depend on the structure of the network Laplacian (and consequently of the adjacency matrix), ruling the current information flow between nodes. To characterize the underlying network substructures, we set out a criterion to scrutinize $\rho$: two nodes can reciprocally process information when they reach a greater than or equal value than the information contained on one of the two nodes, thereby naturally introducing \new{$\rho'_{ij}=\frac{\rho_{ij}}{\min(\rho_{ii},\rho_{jj})}$}. Thus, if two nodes can integrate information depending on their particular $\rho_{ij}$ matrix element at time $\tau$, it is possible to define a new 'information integration' matrix \new{$\zeta_{ij}=\mathcal{H}(\rho'_{ij}-1)$}, where $\mathcal{H}$ stands for the Heaviside step function $\mathcal{H}(x)=\begin{cases} 1~~if~~x\geq0 \\  0~~if~~x<0\end{cases}$. Observe that, for $\tau \to \infty$, the $\rho$ matrix converges to $\rho_{ij}=1/N$, and $\zeta$ is the all-ones matrix, as might be expected.

We have considered the binarized counterpart of the information probability flow, $\zeta$, in terms of the canonical density operator, in analogy to the steepest descent method in the path integral formulation of general diffusion processes \cite{Graham1977}. In particular, following our choice, we are numerically selecting first the most probable paths from  Eq.(\ref{EvolMat}), which gives information about the prominent information flow paths of the network in the interval $0<t<\tau$. 


Figure \ref{Connectome}(a) shows the $P(\rho')$ distribution as a function of $\rho'$ for the weighted HC and different values of $\tau$. By examining $P(\rho')$ for different resolution times of the network, we can conclude that: (i) at short times, no paths are connecting any couple of nodes (i.e., \new{$\rho'_{ij}<1$, $\forall i \neq j$}) and (ii) for large times all values converge to $\rho'=1$, i.e., all the possible paths allow to integrate information between any couple of nodes. In turn, setting neither too big nor too small times enable us to explore the most likely paths of information flow pervading the network structure (i.e., those with $\rho' \geq 1$). Figure \ref{Connectome}(b) summarizes the Laplacian spectrum of the $\zeta$ matrix, for different values of $\tau$. It evolves from a Dirac delta probability distribution, $P(\lambda)=\delta(1/N)$, at time $\tau=0$ (from direct application of Eq.(\ref{EvolMat})) to a progressive convergence to $\lambda_{1}=1$ and $\lambda_i=0$ for $i=2,...,N$, at time $\tau\to\infty$, where $\lambda_1$ correspond to the maximum system's eigenvalue. \new{We also measure the fraction of nodes which are in the giant component, i.e., $GC=\nicefrac{N_{GC}}{N}$. As shown in Fig. \ref{Connectome}(b) the C peak at short times signs the emergence of a non-vanishing giant component percolating troughout the network.} Finally, figure \ref{Connectome}(c) shows the information integration matrix for different diffusion times \new{(evidencing the diffusion mechanism on the different network scales)} together with the network giant connected component (for further examples see also Supplementary Videos \cite{SI}). Also, the singular set of connected nodes emerging from the sharp $C$ criteria, i.e., the information core of the network, is shown in Figure \ref{Connectome}(d)-(g) for different network structures.

\subsection{Synchronizability of the information core}

Synchronization phenomena constitute one of the most glaring examples of dynamical processes where a system needs to properly integrate information among nodes to show an emergent collective state. On the one hand, our results state the existence of the network information core, which stresses the importance of a small group of nodes in integrating information across the network. On the other hand, it is essential to check that, from a dynamical perspective, these nodes play a crucial role in the information processing across the network. This leads us to the critical question: Does the information core generally synchronize for lower values of the coupling strength of the network nodes? Do these nodes dictate the collective behaviour of the network?

The Kuramoto dynamics \cite{kuramoto1975} on a generic network \cite{Arenas2008} is defined by the equation,

\begin{equation}
 \dot \theta_i=\omega_i+K\stackrel[j=1]{N}{\sum}A_{ij}\sin(\theta_j(t)-\theta_i(t))+\sigma \eta_i(t),\label{EqKur}
\end{equation}

where $\theta_i(t)$ represents the phase of the node $i$ at time $t$, $K$ is the coupling strength with all the neighbours, $A_{ij}$ is the adjacency matrix of the network, $\eta_i(t)$ a Gaussian white noise with amplitude $\sigma$, and $\omega_i$ accounts for the intrinsic frequency of each node, being extracted from some arbitrary distribution $g(\omega)$. \new{The Kuramoto order parameter $R(t)=\frac{1}{N}\langle|\sum_je^{i\theta_j(t)}|\rangle$ quantifies the total level of coherence of the system, ranging from $0$ to $1$, where $i$ is the imaginary unit, $|\cdot|$ is the modulus of a complex number, and $\langle \cdot \rangle$ here indicates averages over independent realizations}. Frequency dispersion leads to a critical point at some critical value of $K_c$, separating the synchronous phase from the asynchronous one \cite{Pikovsky,Acebron}.

\begin{figure}[hbtp]
 \includegraphics[width=1.0\columnwidth]{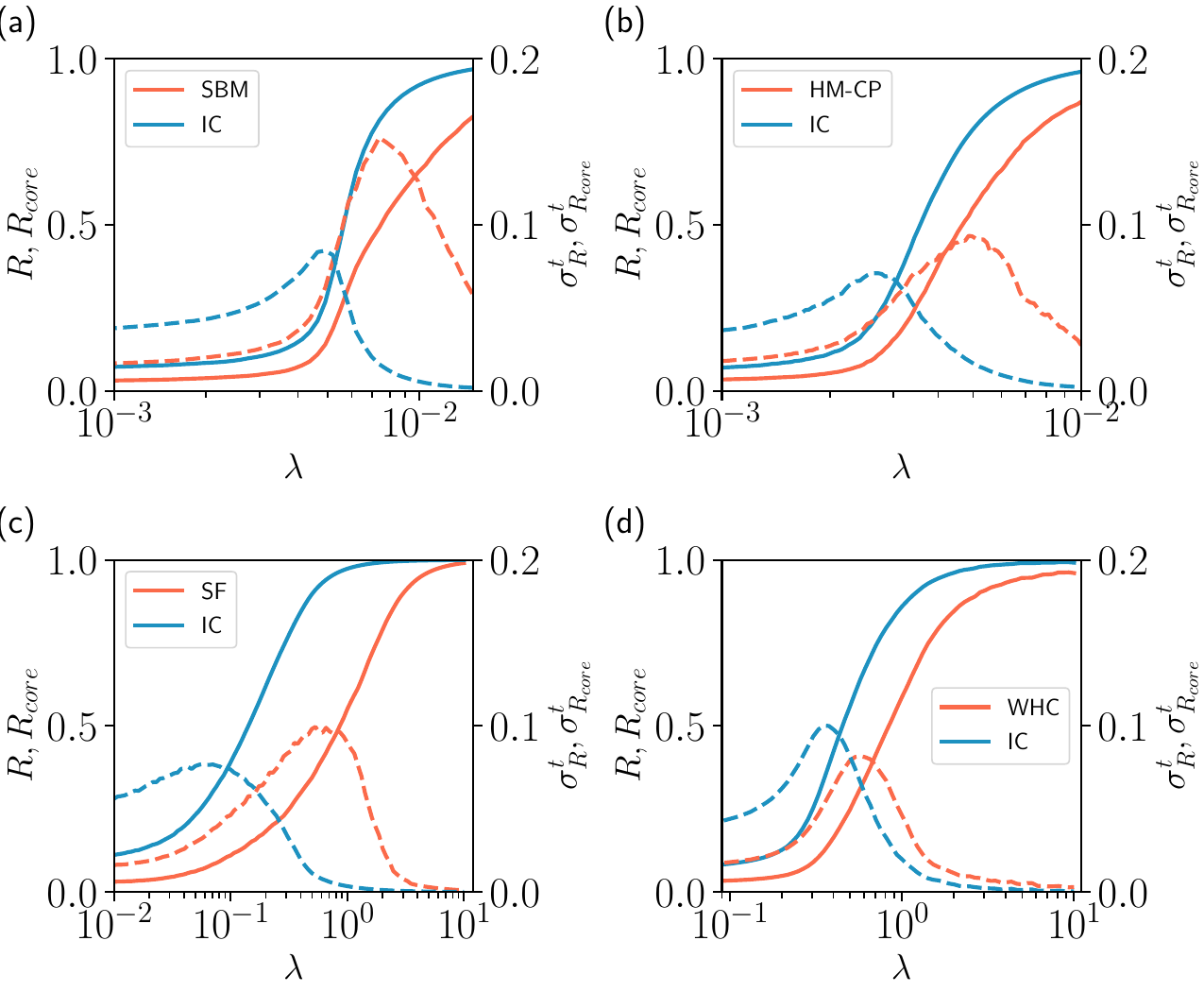}
 \caption{\textbf{Information core synchronizability.} Kuramoto order parameter of the full network and the information core, $R$ and $R_{core}$, versus coupling strength, $K$, for: (a) A SBM of size $N=1024$, with $p=128/N$ and $q=1/N$, (b) a HM-CP network with $N_b=25$, $k_0=12$ and $l=6$,  (c) a SF network of size $N=1024$ and $m=1$ and, (d) the weighted HC. Right y-axis shows the temporal variance of $R$ over network realizations (dashed lines). The point of maximal temporal fluctuations indicate the critical point for the full network and the information core. Observe that the information core generically synchronizes first for all network structures. All curves have been averaged over $10^2$ intrinsic frequency realizations. Parameters: $\sigma=0.1$, $g(\omega)=\mathcal{N}(0,0.1)$, $dt=5\cdot10^{-3}$.}\label{KuramotoC}
\end{figure}

To check our hypothesis, we run computational simulations employing the Kuramoto model \new{with a Gaussian frequency distribution} to test the synchronization efficiency of the information core concerning the whole system. To study in parallel both the global system and the information core of the network, we define two different Kuramoto order parameters capturing the synchronization level either of the global system, $R$, or the giant component of the information core \footnote{In the case of considering multiple, significant, disconnected network components, it is necessary to consider the average over them, $\langle R_{core} \rangle$.}, $R_{core}$.

Figure \ref{KuramotoC} shows the Kuramoto order parameters, $R$ and $R_{core}$, for the relevant cases of different network structures: an SBM, an HM-CP network, an SF network, and the weighted HC. Remarkably, the Kuramoto order parameter and its corresponding temporal variance show that the particular set of nodes conforming to the information core of the network usually exhibits an earlier synchronization phase transition than the entire network (as previously demonstrated, e.g., for trivial SBMs \cite{Villegas1}), even in usual SF networks without a specific hierarchical organization. Thus, we can safely say that the information core generically exhibits coherent behaviour at subcritical collective coupling strength values, managing the integration/segregation of information across the network.

\section{Discussion}

\subsection{Relevant substructures in complex networks} Within our framework, it is possible to define a protocol to identify and analyze the fundamental modules and structures at every appropriate spatio-temporal scale in a complex network, ensuring a sound flux that provides the network connectivity at every scale, indirectly solving thus correlation problems. In particular, the \new{information flow} process from the basal units of the network sheds light on the existence of a deep-meaning substructure, the information core of the network. From the statistical physics perspective, the information core comprises a delicate balance between the internal energy and the intrinsic disorder of the network, shedding light on the key set of nodes controlling the system's dynamical properties. Hence, \new{this 'backbone' in the network} defines an objective and clear criterion to manage complex networks' controllability by altering the dynamical properties of these specific nodes. We have analyzed the information core for different canonical case studies in the field, confirming the expected results for the stochastic block model \cite{Villegas1}, where the information core consists of the basal modules of the network and extracting new essential substructures in scale-free and core-periphery networks. The application to the Human-Connectome network allows us to identify relevant nodes that can be of particular importance in neural functioning and will be analyzed in future works.

Our results confirm previous analyses that showed the importance of core-periphery structures in information processing, i.e., that a central integrative core facilitates the segregation-integration balance optimization \cite{yizhar2011, Gollo2015}. In addition, we have verified the existence of multiple and differentiated scales in the Human Connectome --particularly enhanced by the network weights-- enriching the available dynamical repertoire and the adequate integration/segregation of information of neural networks. This particular structure allows developing sub-modules operating as 'information attractors' --essential for neural functioning \cite{Meunier2010}-- while other structures can manage and distribute information effectively.

\subsection{Dynamical implications and integration-information balance}
In a pioneering work, Tononi and coworkers conjectured the need for an optimal balance between segregation (e.g., several sensory inputs) and integration (allowing for a unified representation, advanced cognitive processing, and response) in the brain for processing high-level cognitive tasks \cite{Tononi}. From a structural viewpoint, we confirm the well-known fact that a hierarchical-modular organization allows for the emergence of an excellent integration/segregation trade-off: segregated information remains trapped in local modules but can travel across the entire network enabling the integration of information between the different modules \cite{Deco2015,Wang2019,Sporns2013}. 

Hence, we hypothesize that the existence of scale-dependent specific structures facilitates information processing across the network. As a direct consequence of this, intrinsically disordered \new{hierarchic-modular} networks do indeed generically optimize transmission and storage of information, improving computational capabilities \cite{Legenstein2007} and strengthening the network stability \cite{Bertschinger2004}. Their particular nested structure has been elucidated to generate, e.g., broad dynamical regions of dynamical criticality, the so-called Griffiths phases, without the need to invoke precise criticality \cite{NatComm}. At the same time, this very structure promoting Griffiths-like phases supports the striking functional variability of synchronization patterns in actual brain dynamics \cite{Tognoli2014, Gollo2019}, also facilitating --from a dynamical perspective-- a flexible balance between segregation and integration et different functional scales \cite{NatComm, Villegas1, Villegas2}.

A very backbone structure controlling the integration/segregation balance over the networks ensures information processing capabilities. For example, the particular application of our method to canonical case studies in the field as, e.g., scale-free networks illustrates the very existence of the short-time peak in $C$, reflecting the presence of a complex core structure, and justifies previous results indicating the existence of a non-vanishing synchronization phase transition \cite{yamir2004}. Still, the entire information transmission throughout the network diverges according to the system size. For example, our approach opens the door to new applications in \emph{network synchronizability}, which depends on how difficult it is to transmit information across the network \cite{Donetti2005}, by only manipulating the dynamical properties of the information core (see \cite{Arenas2008} for a comprehensive review on the topic with further possibilities). 

\section{Conclusions}

Understanding the interrelation of the interspersed structure of physical systems and the scaling laws governing it ({\em }i.e., the problem of pattern and scale) represents a fundamental problem in modern physics \cite{Wilson1979, Levin1992}. To this aim, the statistical physics of phase transitions and, in particular, the RG have been one of the significant developments in contemporary statistical mechanics \cite{Binney, Amit, Kardar}. Their application to diverse dynamical processes operating on top of regular spatial structures (i.e., lattices) allows the introduction of the idea of \emph{universality} and the classification of models (otherwise presumed faraway)  within a small number of universality classes. 

So far, there is no apparent equivalence to analyzing RG processes in complex spatial structures, even if some pioneering approaches have recently proposed clever procedures to state tantamount general RG transformations to those of statistical physics. However, they have always been based on \emph{hidden metric} assumptions, spatially mapping nodes in some abstract topological space, which needs to be considered as an 'a priori' hypothesis \cite{Garcia2018, Zheng2020}. To further advance the issue, it is crucial to develop a field-theoretical framework of complex information dynamics \cite{ghavasieh2020, ghavasieh2022}, based on statistical physics principles, to better understand the mesoscopic interrelationships of complex structures. Hence, exploiting simple diffusion allows to extract information about the network topological space, identify and characterize, e.g., \emph{building blocks} in terms of information diffusion \cite{Boguna2021}, or determine communicability between nodes in the network space. Here, we have taken advantage of the equivalent definition of the canonical density operator \cite{Pathria, Greiner}, which only depends on the Laplacian matrix, governing information diffusion processes in complex networks \cite{Masuda2017}. As a result, it follows the so-called network propagator at time $\tau$, $\hat K (\tau)$, containing all the probabilities (i.e., paths) of broadcast information to neighbour nodes \cite{Delvenne2015}. Thus, we can explore the resolution of the network at multiple scales (making use of the information probability pathways all across the network structure that depend on the diffusion time $\tau$), in an analogous way to the different spatial resolutions as usual done to perform calculations with the RG machinery \cite{Amit}. In particular, we analyze the evolution of information fields through the network's entropy and propose the tantamount of the specific heat, $C$, that reveals 'entropic' phase transitions, a detector of the relevant scales of the system. The specific heat allows us to identify the \emph{information core} of the network, i.e. the information reservoirs where information is firstly stored to be used in other parts of the system. 



We stress that the density matrix \cite{Domenico2016}, $\rho(\tau)$, encodes the time evolution of the information diffusion due to all the elementary paths on the network in a time window $0<t<\tau$. \new{Analogously, the first-passage times of a random walk on top of a complex network may be regarded as a messenger delivering information to each node it visits \cite{Noh2004} (see also \cite{Masuda2017} for further details)}. Therefore, it also \new{characterizes} the probability of remaining trapped in different mesoscale structures of the network and is a proxy to the dynamical trapping previously proposed in widely-used algorithms as InfoMap \cite{Rosvall2007, Rosvall2008} or Markov stability \cite{patelli2020, Barahona2010}. In particular, these two algorithms use specific dynamics strictly related to our approach: the diffusion of a single random walker in the network starting from an arbitrary node to capture specific mesoscopic structures and their hierarchical organizations. Hence, for instance, the length of the trajectories up to trapping can be used to estimate the organization of the network in structures with a fast internal communication but poorly connected among them, thus proposing partitions in modules that require the minimum of bits to be described \cite{Rosvall2008}. It is important to point out that, given the dynamical nature of these algorithms, they use several iterations to neglect fluctuations and, therefore, identify robust communities. The profound relationship with those algorithms will be explored in future works.

Once the different mesoscale characteristic network structures have been investigated, we also can interrelate them with different emergent dynamical properties. For example, we show that generically the information core (that can be seen as a sort of \emph{Matryoshka doll} within the network) is the first substructure to synchronize in the system. In particular, its existence in scale-free networks justifies the non-vanishing synchronization phase transition in these systems. We want to pinpoint the well-known relationship between the topological scales and dynamic time scales in complex networks \cite{Arenas2006}, where transient dynamics towards synchronization strongly depend on the network Laplacian (see \cite{Arenas2006, Villegas1} and Appendix \ref{KuramotoApp}). Nonetheless, while studying such transients can reveal modules in complex networks, they profoundly depend on Taylor series expansions, where non-linear interactions are not considered, and they are always subject to some form of numerical approximation. 

Summing up, our information-based approach constitutes a sound technique for analyzing structures with non-trivial topological features by only considering the diffusion of information centred on the network edges \cite{Masuda2017}. In particular, it can extract the 'topological landscape' of the network at different resolution scales, making fundamental structural blocks emerge and creating a basis for defining crucial blocks in the decimation process of complex networks. Let us finally mention that detailed analyses of the rules allowing for the generation of new supernodes, in the spirit of Kadanoff's blocks, are in progress and will be reported elsewhere. Also, further systematic analyses characterizing possible metrics (e.g., the pioneering work of \cite{Garcia2018}) are still missing, either universal or dependent on general network properties. They can be very illuminating in investigating the spatial projection of complex networks.

Even if further computational and analytical studies would be required to establish a general field-theoretical theory of complex networks definitely, we believe that our approach can open the door to groundbreaking applications for the study of the information flow in the context of gene-regulatory networks \cite{peter2017,erwin2009}, software networks \cite{Fortuna2011, Villegas2020} or protein-protein interaction networks \cite{gagneur2004,kim2014}. Likewise, it represents a significant step forward in developing RG theoretical techniques induced by diffusion distances \cite{Boguna2021}, fostering the definition of supernodes in structures lacking of embedding topological spaces and illuminating scaling laws and multi-scale relationships of complex heterogeneous networks.

\acknowledgments{G.C. acknowledges support from ITALY-ISRAEL project Mac2Mic and EU project nr. 952026 - HumanE-AI-Net. P.V. acknowledge financial support from the Spanish "Ministerio de Ciencia e Innovación” and the "Agencia Estatal de Investigación (AEI)” under Project Ref. PID2020-113681GB-I00. We also thank M.A. Mu\~noz, D. Garlaschelli and V. Buendia for extremely valuable discussions and/or
suggestions on earlier versions of the manuscript.
}

\appendix

\section{Synthetic hierarchical networks} \label{NetBuild}
 
\paragraph*{\textbf{SBM}} The stochastic block model (SBM) is constructed as follows: we define $C_r$ groups of nodes with $n=N/C_r$ nodes on each group, where $N$ is the total number of nodes in the network. Then, they are randomly linked with different intracommunity and intercommunity connection $p$ and $q$, respectively. To ensure the proper scaling --i.e., graphs of constant average degree-- of the network with the system size, we set \new{ $p=\tilde p/N$ and $q=\tilde q/N$}. 

\paragraph*{\textbf{HM-R}} We randomly select two pairs of nodes, connecting them if they were not previously linked, with a dependent probability on their preassigned hierarchical level $l=1,2,...,s$ (as previously proposed in Refs. \cite{NatComm,Zamora2016}).

\paragraph*{\textbf{HM-CP}} In this specific case, connection between modules are not left at random, but with a scale-dependent probability promoting centralized structures between hubs, following the algorithm proposed in \cite{Zamora2016}. We start by creating $2^s$ blocks of $N_s=16$ nodes with mean degree $\kappa_0=12$ at the deepest level. Once this has taken place, we give a weight $p(i)=\nicefrac{i^{-\alpha}}{\sum_j j^{-\alpha}}$, to the $i^{th}$ node of each block, $i=1,2,...,N_s$. Thus, nodes are now taken with probability $p(i)$ and $p(j)$, and connected it they were not already linked.  All the hierarchical levels share the same scale-free exponent $\alpha=2$ except the basal one, with $\alpha=1.7$. It allows us to mimick the empirically supported core-periphery organization with connector hubs in brain structural networks \cite{Eguiluz, Bassett-core, Bassett-core2}.

\section{Entropic phase transitions in SF networks} 
\label{SFAp}

For the sake of completeness, we checked the accuracy of the entropic analysis for scale-free networks created using the Barab\'asi–Albert model.


\begin{figure}[hbtp]
 \begin{center}
 \includegraphics[width=1.0\columnwidth]{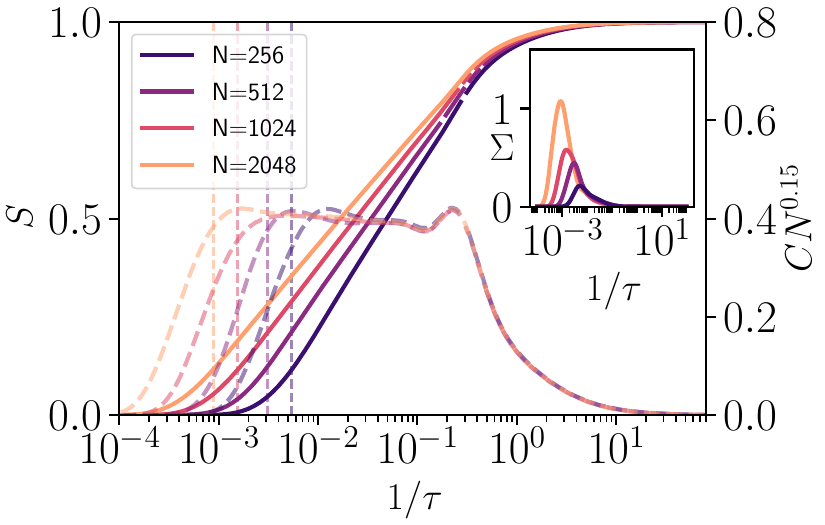}
 \caption{Network average of the entropy parameter $S[\rho(\tau)]$, versus the inverse of diffusion time $\nicefrac{1}{\tau}$, for a Barabasi-Albert network \new{with $m=1$, $\langle k \rangle=2$} and different sizes (see legend). The critical point ($\tau_C$) separates the segregated phase from the integrated one. The specific heat, $C$, scales depending on the system size at this critical value, even if it presents a local peak at short diffusion times. Inset: Variance of the entropy averaged over network realizations multiplied by $N$; $\Sigma=\sigma_S N$. The point of maximal variability mark the full integration of information throughout the network for all network sizes $N$ (dashed lines in main figure). All curves have been averaged over $10^2$ realizations.}\label{SFree}
 \end{center}
\end{figure}

Here, we investigate whether scale-free networks present some of the above described characteristic structures at some network scale. Results of our computational analyses are reported in Figure \ref{SFree}, which displays the entropic order parameter, $S$, at different temporal resolution scales, $\tau$, for multiple system sizes (see legend). The SF networks present a vanishing phase transition for large evolution times (see the entropic scaled variance, $\Sigma$, in the inset of Fig.\ref{SFree}), even if a local peak of $C$ for short evolution times, whose position does not depend on the system size, justifies the existence of the core structure controlling the information processing all across the network, as previously demonstrated.






\section{Core-periphery structural effects} \label{CorePAp}
To gain analytical insight into the effects of sound topological structures as core-periphery, here we analyze two analogous hierarchical networks (with an equal number of nodes, basal modules, and the total number of hierarchical levels) but including core-periphery effects in one of the networks \cite{Zamora2016}.

\begin{figure}[hbtp]
 \begin{center}
 \includegraphics[width=0.95\columnwidth]{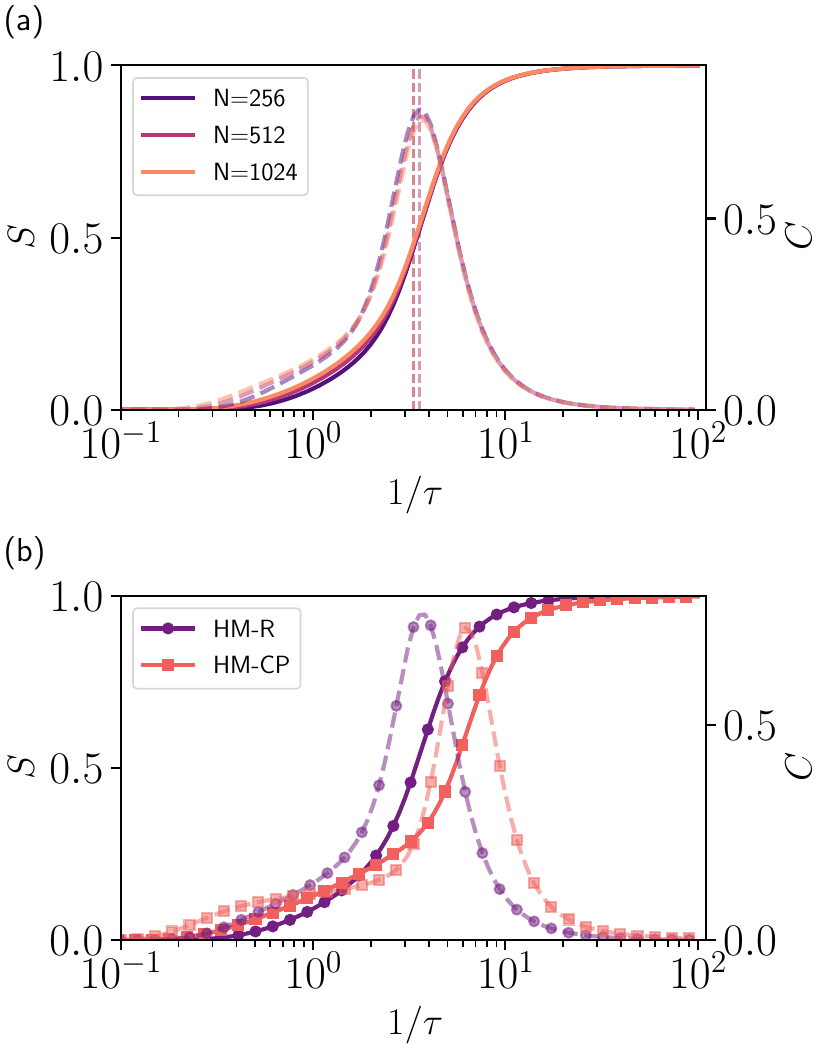}
 \caption{Network average of the entropy parameter $S[\rho(\tau)]$, versus the inverse of diffusion time $\nicefrac{1}{\tau}$ for: (a) \new{HM-R} network for different network sizes, $N=2^s$ (see legend), where $s$ is the total number of hierarchical levels. The slow decay in the derivative of the order parameter, $C$, confirms the existence of a broad region separating the segregated phase from the integrated one. (b) Comparison between a hierarchic modular network with core-periphery structure (HM-CP, \new{$\langle k \rangle\simeq47$}) and a simple hierarchic modular network (HM-R, \new{$\langle k \rangle\simeq28$}) with $6$ hierarchical levels. The HM-CP network exhibits shorter integration times and a broader regime of information processing in the zooming out process. All curves for the HM-R and the HM-CP have been averaged over $10^2$ realizations.}\label{CPer}
 \end{center}
\end{figure}

Figure \ref{CPer} (a) shows the entropic phase transition for an \new{HM-R} network, increasing its total number of hierarchical scales. A broad region emerges where the different spatial scales aggregate when the zooming out process is considered. However, as shown in Figure \ref{CPer} (b), the very existence of a core-periphery structure allows to a more efficient information processing (at shorter times, i.e., resolution scales) and to broaden the different characteristic scales of the network, facilitating information to remain trapped into characteristic scales or modules. In conclusion, core-periphery structures enable broader information processing possibilities than regular hierarchic modular networks.

\section{Structural effects on Kuramoto dynamics} \label{KuramotoApp}

The particular case of the Kuramoto model with no noise ($\sigma=0$) and all identical frequencies (e.g., $\omega_i=0$) allow us to focus specifically on structural effects \cite{Villegas1, Arenas2008}. Due to the absence of noise,  large populations with no frequency dispersion always reach the overall synchronized state ($R=1$). In particular, assuming that, at some significant time $t$, the phases will be sufficiently small, it is possible to consider the Taylor series expansion of Eq. \ref{EqKur}, which reads, 

\begin{equation}
 \dot \theta_i=K\left[\underset{j}{\sum}A_{ij}\theta_{j}-\underset{j}{\sum}\delta_{ij}\left(\underset{l}{\sum}A_{jl}\right)\theta_{j}\right]=-K\underset{j}{\sum}L_{ij}\theta_{j},
\end{equation}

where $L$ represents the Laplacian matrix of the network. It allows, for example, modular identification techniques based on routes towards synchronization, like the one proposed by Arenas et al. \cite{Arenas2006}. Nevertheless, as discussed above, we stress that these approaches always depend on numerical methods and rest upon a Taylor series expansion.

\newpage

%

\end{document}